\documentclass[12pt]{iopart}

\usepackage{graphicx}% Include figure files
\usepackage{dcolumn}% Align table columns on decimal point
\usepackage{bm}% bold math
\usepackage{amssymb}
\begin{document}

\title{Measurement Induced Quantum Coherence Recovery}

\author{JinShi Xu, ChuanFeng Li, Ming Gong, XuBo Zou, Lei Chen, Geng Chen, JianShun Tang and GuangCan Guo}

\address{Key Laboratory of Quantum Information,
  University of Science and Technology
  of China, CAS, Hefei, 230026, People's Republic of China}
\ead{cfli@ustc.edu.cn}
\begin{abstract}
We show that measurement can recover the quantum coherence of a
qubit in a non-Markovian environment. The experimental demonstration
in an optical system is provided by comparing the visibilities (and
fidelities) of the final states with and without measurement. This
method can be extended to other two-level quantum systems and
entangled states in a non-Markovian evolution environment. It may
also be used to implement other quantum information processing.

\end{abstract}

\maketitle

\section{Introduction}
At the beginning of the development of quantum mechanics, measuring
problem is treated with the project measurement model given by Von
Neumann \cite{Neumann55}. In this model, measuring process leads to
the irrevocable collapse of the quantum system into eigenstates and
the coherence is destroyed. During the last decades, with the
development of quantum information theory, quantum measurement has
been understood in the framework of quantum decoherence theory
\cite{Wheeler83} and has been used to construct some quantum
information processes, such as the Knill, Laflamme, and Milburn
(KLM) scheme \cite{Knill01} and one way quantum computation
\cite{Raussendorf01}, etc. Specially, quantum Zeno effect with
continuous measurement can be used to preserve the coherence of
specific states \cite{Misra77,Nakanishi01}. Recently, it has been
shown that weak measurement can erase the collapse effect induced by
a previous weak measurement and the initial quantum state can be
recovered \cite{Korotkov06}. Katz \etal have experimentally verified
this idea using superconducting phase qubits \cite{Katz08}.

In this paper, we show that measurement can recover quantum
coherence of a single qubit evolved in a non-Markovian environment
which has the memory effect. A theoretical description of this
method which predicts that the visibility of the qubit can be
recovered from 0 to 50\% is given. Then we demonstrate this
phenomenon experimentally in an optical system with photons produced
by the process of spontaneous parametric down-conversion. The
recovery can be seen clearly by comparing the visibilities (and
fidelities) of the final states with and without measurement. We
also provide a simplified picture to understand this phenomenon.

\section{Theoretical description}
The polarization of a single photon is used as the information
carrier and birefringent elements, which can couple the photon's
frequency with its polarization, are adopted as the adjustable
non-Markovian environment. Consider an arbitrary input pure
polarization state
\begin{equation}
|\psi\rangle=\alpha|H\rangle+\beta|V\rangle,
\end{equation}
where $\alpha$ and $\beta$ are complex numbers which obey
$|\alpha|^{2}+|\beta|^{2}=1$. $|H\rangle$ and $|V\rangle$ represent
the horizontal and vertical polarization states, respectively. As a
result, the output after interaction time $t$ in the birefringent
crystal can be written as
\begin{equation}
|\psi(\omega,t)\rangle=\alpha|H\rangle+e^{i\kappa\omega
t}\beta|V\rangle,
\end{equation}
when the optic axis of the birefringent crystal is set to be
horizontal. The parameter $\kappa$ is proportional to $\Delta
n=n_{o}-n_{e}$, which is the difference between the indexes of
refraction of ordinary ($n_{o}$) and extraordinary ($n_{e}$) light.
Because $\Delta n\ne 0$, different frequency will introduce
different phase shift $\kappa\omega t$ in the output state.
Considering the contributions of all the frequencies, the relative
phase between the information carrier bases $|H\rangle$ and
$|V\rangle$ may become truly uncorrelated, which will destroy the
coherence of the qubit \cite{Kwiat00}.

This phenomenon is quite similar to the Rabi oscillation of a qubit
in an external field. For a Rydberg atom in a cavity
\cite{Auffeves03}, the overall Rabi oscillation should be integrated
over the photon number distribution which corresponds to the
frequency distribution $f(\omega)$ of the photon in our case.
Therefore, the polarization state of the photon can be written as
the following reduced density operator \cite{Berglund00}
\begin{equation}
\rho=\int f(\omega)|\psi(\omega, t)\rangle\langle \psi(\omega, t)|
\mathrm{d}\omega. \label{eq:rho0}
\end{equation}
The final total probability for us to detect $|\psi\rangle$ after
interaction time $t$ is
\begin{equation}
\mathcal{P}_{|\psi\rangle}=|\alpha|^{4}+|\beta|^{4}+2|\alpha|^{2}|\beta|^{2}\int
f(\omega)\cos(\kappa\omega t)\mathrm{d}\omega. \label{eq:eq0}
\end{equation}
$\mathcal{P}_{|\psi\rangle}$ tends to $|\alpha|^{4}+|\beta|^{4}$,
when the interaction time is sufficiently large \footnote[1]{For the
continuous frequency distribution $f(\omega)$, $\lim_{t\rightarrow
\infty} \int f(\omega) \exp(i\omega t)d\omega=0$.}. It can be seen
that when $|\beta|^{2}=\frac{1}{2}$ the fidelity tends to $1/2$ and
the visibility
$\mathcal{V}_{|\psi\rangle}=2\mathcal{P}_{|\psi\rangle}-1\rightarrow0$,
which means that the qubit loses its coherence completely
\cite{Berglund00}. However, the coherence will be restored by
measuring the qubit during its evolution, as can be seen below.

We insert measurement apparatus with the measurement basis setting
at $\pm45^{\circ}$ polarization ($+/-$), where the two projected
states are separated into two paths 1 and 2 without disturbing the
photon's subsequent dynamics. It should be noticed that the
frequency distributions of the projected states after measurement
are different from that of the initial state and the Pauli
$\sigma_{x}$ operation is employed to reverse $H$ and $V$ of the
final output state. If the interaction time is $t$ before
measurement while the residual interaction time is $t'$ after
measurement, the output density operator of the polarization reads
as (the subscripts represent the paths 1 and 2)
\begin{equation}
\rho'=\int f(\omega) (K_{+}|\varphi\rangle_{1}\langle \varphi| +
K_{-}|\varphi\rangle_{2}\langle \varphi|)\mathrm{d}\omega,
\label{eq:eq1}
\end{equation}
where $K_{+}=\frac{1}{2}(1+2|\alpha||\beta|\cos(\phi+\kappa\omega
t))$ and $K_{-}=\frac{1}{2}(1-2|\alpha||\beta|\cos(\phi+\kappa\omega
t))$ are the probabilities of projecting into $+$ and $-$
polarization states, respectively. $\phi$ is the relative phase
between $|H\rangle$ and $|V\rangle$ of the initial state.
$|\varphi\rangle_{1}=\frac{1}{\sqrt{2}}(|V\rangle_{1}+e^{i\kappa\omega
t'}|H\rangle_{1}$ and
$|\varphi\rangle_{2}=\frac{1}{\sqrt{2}}(|V\rangle_{2}-e^{i\kappa\omega
t'}|H\rangle_{2})$ are the residual evolution states in paths 1 and
2, respectively. As a result, the total probability for us to find
$|\psi\rangle$ is
\begin{equation}
\mathcal{P}^{'}_{|\psi\rangle}=\frac{1}{2}+2|\alpha|^{2}|\beta|^{2}\int
f(\omega)\cos(\phi+\kappa\omega t)\cos(\phi+\kappa\omega
t')\mathrm{d}\omega. \label{eq:3}
\end{equation}
At the limit of long enough interaction time and with $t'=t$,
$\mathcal{P}^{'}_{|\psi\rangle}$ tends to
$\frac{1}{2}+|\alpha|^{2}|\beta|^{2}$ and we can get coherence
recovery for a set of pure states, which can be seen from figure 1.
At the area
$\frac{1}{6}(3-\sqrt{3})<|\beta|^2<\frac{1}{6}(3+\sqrt{3})$ the
fidelity with measurement is larger than the one without measurement
and it reaches its maximal recovery at $|\beta|^{2}=\frac{1}{2}$
where the fidelity restores to 75\%. The states which close to the
eigenstates of the decoherence environment $H/V$ are less decohered
and measuring the qubit can not improve the fidelity.  For the set
of maximally recovered states with the form
$\frac{1}{\sqrt{2}}(|H\rangle+e^{i\phi}|V\rangle)$, the recovered
fidelity 75\% is larger than the classical limit 66.7\%
\footnote[2]{The projective probability of any orthorgnal
measurement basis of the subspace spanned by these maximally
recovered states distributes on [0,1]. Therefore the fidelity
allowed by classical optics is 66.7\%, see \cite{Preskill98}.},
which shows their quantum effect.

\begin{figure}[tbph]
 \begin{center}
 \includegraphics[width= 3.in]{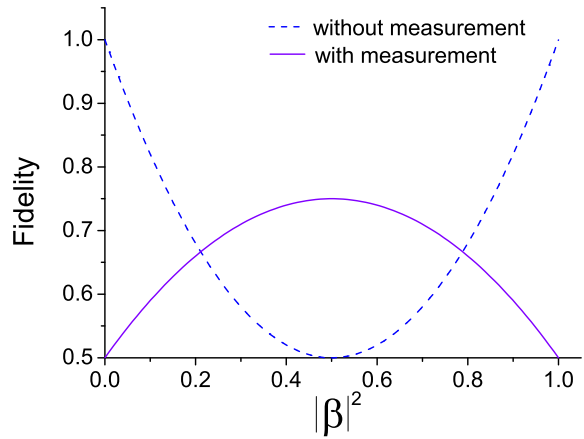}
 \end{center}
 \caption{The fidelity at different cases with the interaction time long enough and $t=t'$
 for the case with measurement during the evolution.}
 \end{figure}

\section{Experimental demonstration and discussion}

In order to experimentally demonstrate this phenomenon, we choose
the initial state
$|+\rangle=\frac{1}{\sqrt{2}}(|H\rangle+|V\rangle)$ from the set of
maximally recovered states. The setup of the experiment is shown
schematically in figure \ref{fig:setup}. The second harmonic
ultraviolet (UV) pulses are frequency doubled from a mode-locked
Ti:sapphire laser with the center wavelength mode locked to 0.78
$\mu$m (with 130 fs pulse width and 76 MHz repetition rate). These
UV pulses are focused into a beamlike cut beta-barium-borate (BBO)
crystal \cite{Weinfurter00,Takeuchi01} to produce highly bright
spontaneous parametric down-conversion (SPDC) photons with special
polarizations. We get about 28000 coincidence events per second and
the integral time is 10 s for each measurement. One of the SPDC
photons (path $b$) is prepared into $|+\rangle$ to demonstrate the
coherence recovery while the other (path $a$) is used as a trigger.

\begin{figure}[tbph]
 \begin{center}
 \includegraphics[width= 3.in]{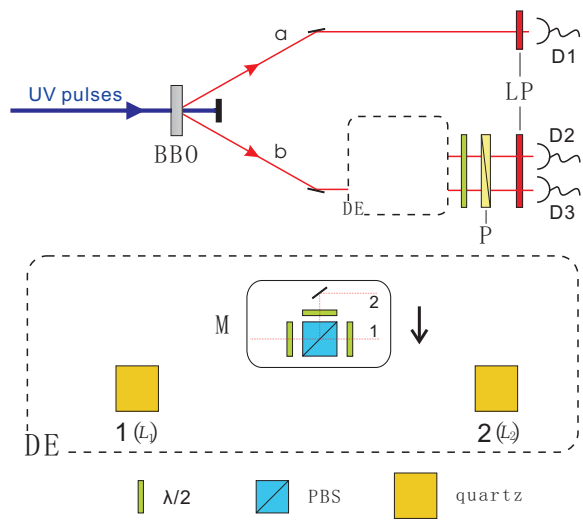}
 \label{fig:setup}
 \end{center}
 \caption{Experimental setup. The decoherence evolution DE is
 denoted by a dashed pane. The measuring apparatus (M) is inserted
 depending on different cases. We use half-wave plates ($\frac{\lambda}{2}$) to reverse $H$ and $V$ of the output state.
 The final detecting bases are chosen by a polarizer (P). Long pass filters (LP) are equiped in front of
 single-photon detectors to minimize the influence of the pump light. Any
successful detection is given by the coincidence of single-photon
detectors D1 and D2 (path 2) or D1 and D3 (path 1).}
\label{fig:setup}
 \end{figure}

The decoherence evolution of the signal photon in path $b$ is the
controllable birefringent ``environment" using quartz plates with
thickness $L$ which are distributed into two sets (set 1 with
thickness $L_{1}$ and set 2 with thickness $L_{2}$). Measurement
apparatus (M), which contains three half-wave plates ($\lambda /2$)
with optic axes setting at the same angle of $22.5^\circ$ according
to the axis of quartz and a polarization beam splitter (PBS), can
project a photon state onto $+$ or $-$ linear polarization which is
corresponding to the path 1 or 2, respectively. We use a polarizer
(P) in path $b$ to choose the final detecting polarization of the
signal photon. Both photons are then coupled by multi-mode fibers to
single-photon avalanche detectors which are equipped with long pass
filters (LP) to minimize the influence of the pump light. Any
successful detection is given by the coincidence of the trigger
photon (D1) and the concerned photon (D2 or D3).

The frequency spectrum of the photon is considered as a Gaussian
amplitude function $G(\omega)$ with frequency spread $\sigma$ and it
is peaked at the central frequency $\omega_{\circ}$ corresponding to
the central wavelength $\lambda_{\circ}=0.78$ $\mu$m.
\footnote[3]{Deduced from the Gauss-like pulse pumping laser.}
According to equation (\ref{eq:eq0}) which is the case without
measurement, the total probability of detecting $|+\rangle$ is
\begin{equation}
\mathcal{P_{+}}=\frac{1}{2}+\frac{1}{2}\cos(\gamma\omega_{\circ})e^{-\gamma^{2}\sigma^{2}/16},
\label{eq:before}
\end{equation}
where $\gamma=L\Delta n/c$ and $c$ represents the velocity of the
photon in the vacuum. In our experiment, we can treat $\Delta n$ as
a constant of 0.01 for the small frequency distribution and the
thickness of quartz plates $L$ is represented by the corresponding
retardation $x$, which obeys the equation $L=x/\Delta n$. The
visibility of the final state without measurement can be calculated
as
$\mathcal{V}_{+}=\cos(\gamma\omega_{\circ})e^{-\gamma^{2}\sigma^{2}/16}$.
We can see that it will tend to zero with the increasing of the
thickness of quartz crystals.

However, if we measure the photon by inserting M between $L_{1}$ and
$L_{2}$, we can obtain certain coherence recovery. According to
equation (\ref{eq:3}), we get the final total probability of
detecting $|+\rangle$ (for $L>L_{1}$)
\begin{equation}
\mathcal{P^{'}_{+}}=\frac{1}{4}(2+\cos(\gamma\omega_{\circ})e^{-\gamma^{2}\sigma^{2}/16}+\cos(\xi\omega_{\circ})e^{-\xi^{2}\sigma^{2}/16}),
\label{eq:after}
\end{equation}
where $\xi=(2L_{1}-L)\Delta n/c$.

It can be seen that for large $L$ and $L_{1}=L_{2}=L/2$, we still
have the probability of $0.75$ to detect $|+\rangle$ compared to
$0.5$ in the case without measurement. The visibility in this case
is
$\mathcal{V}^{'}_{+}=\frac{1}{2}+\frac{1}{2}\cos(\gamma\omega_{\circ})e^{-\gamma^{2}\sigma^{2}/16}$,
which can finally tend to 0.5 with the increasing of $L$.

\begin{figure}[tbph]
\begin{center}
\includegraphics[width= 2.9in]{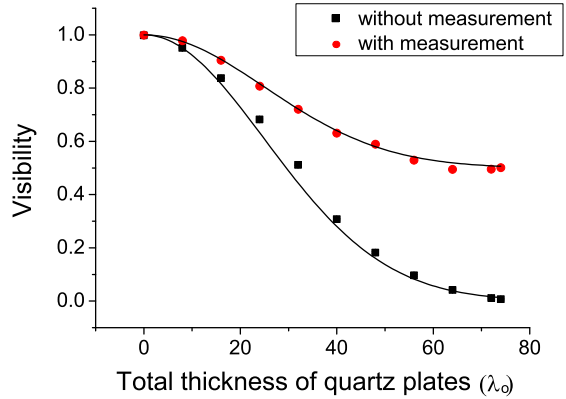}
\end{center}
\caption{Experimental results for the visibility in different cases.
The solid lines are the theoretical results. The thickness of quartz
plates is represented by the retardation. $\lambda_{\circ}=0.78$
$\mu$m. The error bars which due to counting statistics are less
than the size of the symbols.} \label{fig:visibility}
\end{figure}

The visibility of the final state as a function of thickness $L$ is
presented in figure \ref{fig:visibility}, where the dots represent
the data employing measurement during the decoherence evolution and
the squares represent the data without measurement. For each
thickness $L$ we let $L_{1}=L_{2}=L/2$ to get the corresponding
visibility in the case with measurement. We tilt the quartz plates
with optic axes set to horizontal so that the relative phase is the
integral multiple of $360^\circ$ \cite{Kwiat00}. The solid lines are
the theoretical fittings using the equations $\mathcal{V}_{+}$
(without measurement) and $\mathcal{V}^{'}_{+}$ (with measurment)
mentioned above. In our experiment, the frequency spread is about
$6.9\times10^{12}$ Hz. It is shown that when the total thickness of
quartz plates is increased to 74$\lambda_{\circ}$, the visibility
with measurement reaches 0.501 while it will drop close to zero
without measurement. Good fittings between theoretical predictions
and experimental data are found.

\begin{figure}[tbph]
\begin{center}
\includegraphics[width= 2.9in]{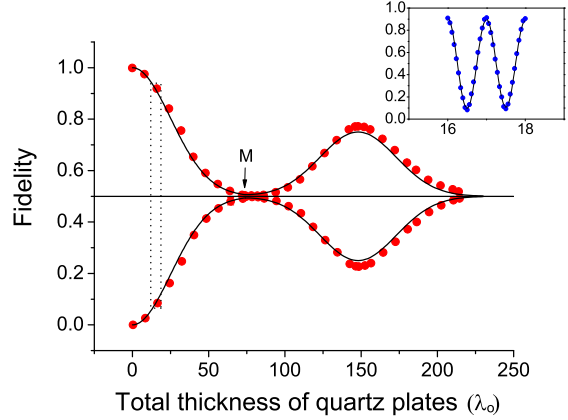}
\end{center}
\caption{Experimental results for the detecting fidelity.
Measurement apparatus (M) and $L_{2}$ are inserted (denoted by the
arrow) when $L_{1}$ reaches 74$\lambda_{\circ}$. The solid lines are
the theoretical results employing equation (\ref{eq:before}) (before
measurement) and equation (\ref{eq:after}) (after measurement). The
inset is the oscillation between maximal and minimal result in the
dotted pane.} \label{fig:fidelity}
\end{figure}

We further demonstrate this phenomenon in a visualized way by
measuring the fidelity of the state in the whole evolution. As shown
in figure 4, we insert measurement apparatus (M) and $L_{2}$ when
$L_{1}$ increases to 74$\lambda_{\circ}$. While $L_{2}$ increases to
74$\lambda_{\circ}$ too, we obtain the highest probability 0.773 to
get $|+\rangle$ corresponding to the theory prediction of 0.75. This
error is mainly due to the limitation of precision when we calibrate
the axes of the quartz plates. We also show the oscillation between
the maximal and minimal probability of getting $|+\rangle$ by
tilting a quartz plate to get the required angles. It can be seen
from the inset in figure 4 that the oscillation is similar to a
cosine curve in a small distribution of $L$ which agrees well with
the theory prediction. As a result, we have experimentally
demonstrated the coherence recovery by measuring the photon in the
evolution.

\begin{figure}[tbph]
 \begin{center}
 \includegraphics[width= 2.9in]{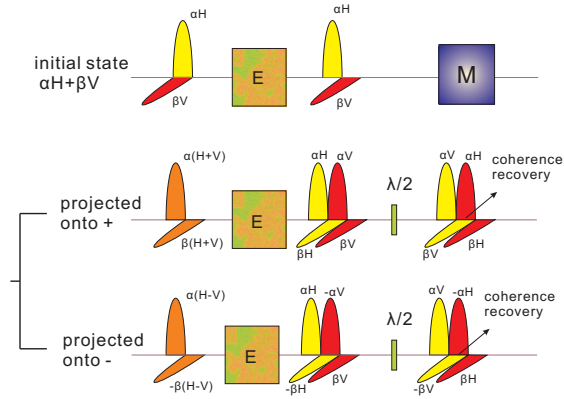}
 \end{center}
 \caption{Simplified picture to understand the
 coherence recovery phenomenon. All the sates at different evolution
 times are represented by wave packets with special polarizations. E is the
 decoherence evolution and M represents the measurement apparatus which projects the photon onto $+$ or $-$ polarization.} \label{fig:understand}
 \end{figure}

As measurement preserves the frequency distributions and the
relative phases of the two projected states in the non-Markovian
environment, we can choose a suitable interaction time after
measurement to erase some of the unwanted effects of the
environmental interaction. We may understand this phenomenon in a
simplified picture as shown in figure \ref{fig:understand} in time
domain. The coherent superposition of the initial state of the
photon comes from the overlap in temporal modes of the two
eigenstates $|H\rangle$ and $|V\rangle$. The first set of quartz
plates with enough thickness which represented by E in figure
\ref{fig:understand} destroys the overlap completely. The projected
states in the basis $+/-$ after measurement will preserve their
relative phases. After passing through the second set of quartz
plates with the same thickness some of the eigenstate components
will overlap and then get the coherence recovery. We should only
insert a half-wave plate to transfer the state of recovered part
into the initial state acting as a $\sigma_{x}$ operation. Figure
\ref{fig:understand} also implies that it is possible to get perfect
coherence recovery by employing the time bin technology
\cite{Brendel99} to select only the recovered part.

\section{Conclusion}
In conclusion, we have demonstrated that by measuring a photon qubit
during its evolution in a non Markovian environment, the destroyed
coherence can be recovered. It can be deduced from the theoretical
mode we give that this kind of measurement may be also realized in
other two-level systems such as a Rydberg atom coupled to a
microcavity \cite{Haroche05} and an electronic spin coupled to
nuclear spins \cite{Yao07, Witzel05}. This technology is also useful
to demonstrate entanglement recovering, Leggett-Garg inequality
\cite{Legg85} and some kinds of Bell-like inequalities
\cite{Zela07}.

\section*{Acknowledgments}
This work was supported by National Fundamental Research Program,
the Innovation funds from Chinese Academy of Sciences, National
Natural Science Foundation of China (Grant No.60121503) and Chinese
Academy of Sciences International Partnership Project.

\end{document}